\documentclass[prl,twocolumn,preprintnumbers,amsmath,amssymb,floatfix,reprint]{revtex4}
\usepackage{hyperref}
\usepackage{epsfig}
\begin{document}

\title{Sliding Over a Phase Transition}

\author{A. Benassi$^{1,2}$, A. Vanossi$^{2,1}$, G.E. Santoro$^{2,3,1}$ and E. Tosatti$^{2,3,1}$}
\affiliation{
$^1$ CNR-IOM Democritos National Simulation Center,Via Bonomea 265, I-34136 Trieste, Italy \\
$^2$ International School for Advanced Studies (SISSA), Via Bonomea 265, I-34136 Trieste, Italy \\
$^3$ International Centre for Theoretical Physics (ICTP), P.O.Box 586, I-34014 Trieste, Italy
}

\date{\today}


\begin{abstract}
The effects of a displacive structural phase transition on sliding friction are in principle accessible to nanoscale
tools such as the Atomic Force Microscopy, yet they are still surprisingly unexplored. We present model simulations
demonstrating and clarifying the mechanism and potential impact of these effects. A structural order parameter
inside the material will yield a contribution to stick-slip friction that is nonmonotonic as temperature crosses
the phase transition, peaking at the critical $T_c$ where critical fluctuations are strongest, and the sliding-induced
order parameter local flips from one value to another more numerous. Accordingly, the friction below $T_c$
is larger when the order parameter orientation is such that flips are more effectively triggered by the slider.
The observability of these effects and their use for friction control are discussed, for future application to
sliding on the surface of and ferro- or antiferro-distortive materials.
\end{abstract}

\maketitle

Understanding and controlling nanoscale friction are among the top priorities in nanoscience and technology, where
moving elements are increasingly important. Unearthing mechanisms capable of altering dry friction between solids,
to be employed in addition to traditional means such as lubrication, tuning of load, temperature and speed~\cite{persson_book},
is of great interest, in physics as well as for potential applications. The idea is to replace a ``dead'' substrate,
whose role is purely passive, with a ``live'' material hosting a phase transition. Early surface science work provided an
indirect hunch, in the form of a predicted drop of the two-dimensional diffusion coefficient $D$ for a brownian adsorbate particle,
caused by an underlying surface phase transition~\cite{ying,ying2,prestipino}. In that case, Einstein's relation
$D \eta = k_BT$ implies the prediction of a peak in the particle's viscous friction $\eta$ at the surface critical
temperature $T^{surf}_c$. Although this linear response mechanism is not realistically relevant to dry friction,
which is dominated by stick-slip and strong nonlinearities~\cite{persson_book}, that example is nonetheless
suggestive of a frictional anomaly near a bulk phase transition of the substrate. Experimentally, a spectacular anomaly,
elegant albeit confined to a very special case of electronic friction, is the critical drop of frictional dissipation (noncontact and contact) reported
at the superconducting-normal transition of a metal substrate~\cite{kisiel,krim}.

In some ferroelectrics such as TGS (triglycine sulfate)~\cite{correia} and to a lesser extent BaTiO3~\cite{eng2}, AFM topography and
friction have shown contrast between surface domains. However, the critical temperature dependence of sliding
friction has not been explored; and these systems are pestered with structural and electrostatic complications
which one may wish to avoid at this first, more fundamental level. At that level, the basic questions are
(i) what is the frictional coupling mechanism between tip motion and a substrate structural phase transition;
(ii) what is the distinguishing element of the phase-transition related frictional contribution relative to the background friction; and
(iii) could one hope to achieve some friction control through external fields that influence the substrate order parameter?

\begin{figure}
\centering
\includegraphics[width=8.5cm,angle=0]{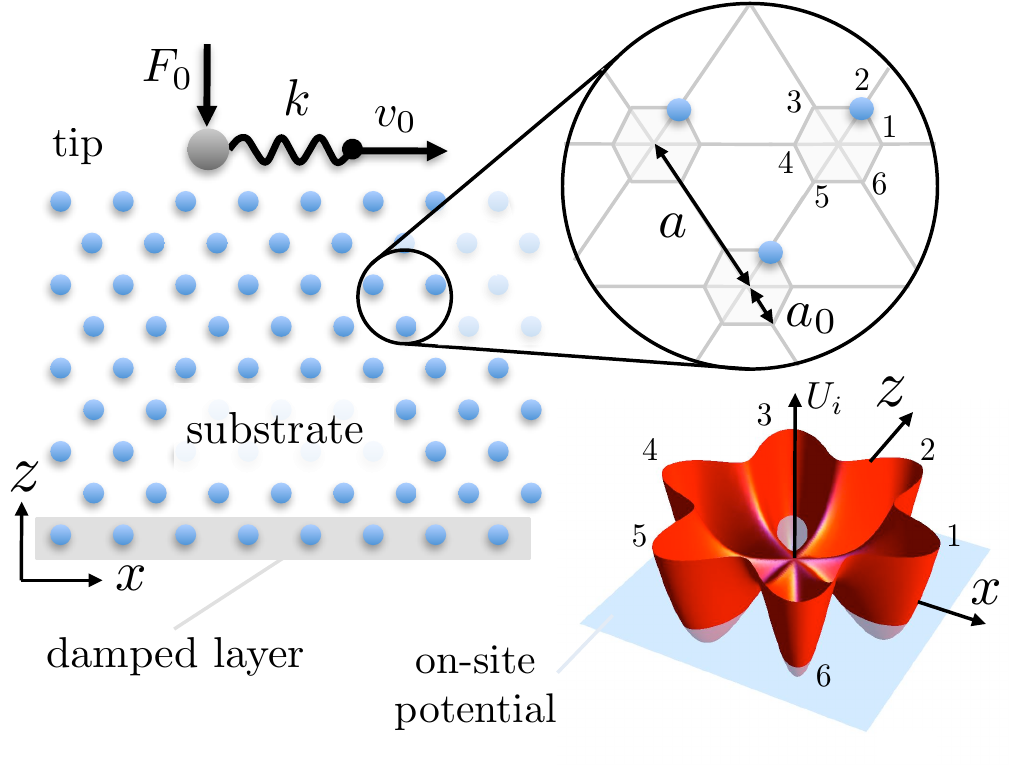}
\caption{Sketch of the 2D model system, the zoom shows the on-site potential symmetries, with the
hexagon vertices representing the six valleys of the on-site potential, displayed below.}
\label{figura1}
\end{figure}

In order to explore these questions theoretically ahead of future experiments, we resort to a model study.
Specifically, we carry out Molecular Dynamics (MD) classical simulations for the stick-slip dry friction of an
idealized point-like AFM tip over a two-dimensional (2D) model crystal substrate whose bulk undergoes a
weak, continuous structural phase transition. Our chosen model will be one of a general ``displacive'' type, a
category in principle describing popular systems such as some ferroelectrics~\cite{cowley, lines_glass}.


Fig.~\ref{figura1} sketches our model system, a 2D close-packed atomic lattice (the substrate) of classical particles
(the atoms), and a point slider (the tip) pulled over the substrate edge (the surface) through a spring (the cantilever), as in
a classic Tomlinson model~\cite{persson_book}. The substrate atoms (of mass $m$) are held together at
an average distance $a$ by an interatomic pair potential $U_{ij}$, locally similar to a Lennard Jones (LJ) potential,
$U_{ij}=-U+\alpha(\vert \mathbf{r}_i-\mathbf{r}_j\vert-a)^2+\beta(\vert \mathbf{r}_i-\mathbf{r}_j\vert-a)^4$,
an approximation which avoids complications including thermal expansion.
Parameters $\alpha$ and $\beta$ are obtained by fitting a LJ potential of depth $U$ and radius $a$ ($U$, $a$ and
the mass $m$, define our set of ``natural'' units). In an interval $[-0.1,0.1]$ centered on the minimum at $a$,
the fit yields $\alpha=28.32 U/a^2$ and $\beta=784.35 U/a^4$. In addition to the interatomic interaction $U_{ij}$,
each atom is subject to a six-valley on-site potential $U_{i}$ with the same symmetry as the lattice
(see inset in Fig.~\ref{figura1}) chosen such as to favor a small uniform distortion of all sites towards one
of six equivalent valleys $\lambda=1,...,6$
 $U_{i} = U_{M}-\frac{2(U_{M}-U_{m})}{a_0^2}\bigg(3\frac{x_i}{{u}_i}-4\frac{x_i^3}{{u}_i^3}\bigg){u}_i^2 + \frac{U_{M}-U_{m}}{a_0^4}{u}_i^4$.
Here ${u}_i^2=x_i^2+z_i^2$ is the displacement magnitude of the $i$-th atom from the site center, $a_0$ is the distance
between the minima and the center, and $U_{M}$ and $U_{m}$ are the height and the depth of maximum and minima
($U_M=0.1 U$, $U_m=-0.1 U$ and $a_0=0.05 a$). At low temperature $T \ll \vert U_M - U_m\vert$, the ideal 2D
lattice is characterized by the displacive vector order parameter
${\boldsymbol \delta}(T) = \langle{\mathbf u}\rangle = \langle {\mathbf r} -{\mathbf r_{0}} \rangle$,
measuring the average distortion $\mathbf{u}$ of atoms from the central triangular lattice positions
${\mathbf r}_{0}$. At $T=0$, the substrate minimal energy state is at $\vert {\boldsymbol \delta}(0)\vert=a_0$,
with all the atoms in the same valley $\lambda$, and $a_0/ a \ll 1$.
MD simulations of this model are carried out using 2D rectangular cells of large but finite
thickness $L_z = N_z a \sqrt{3}/2$ along z and length $L_x =N_x a$ (typically $N_x=40, N_z =40$),
first in thermal equilibrium, with bulk-like periodic boundary conditions (PBC)
applied along both $x$ and $z$; then out of equilibrium, with PBC along $x$ alone, the
tip sliding taking place on one of two edges, in frictional simulations.
We stress here our intent to mimic qualitatively the behaviour of a real 3D system with a continuous
phase transition, with the order parameter coupled to the slider's motion ~\footnote{
Fine details of the substrate critical behavior are purposely ignored here . Thus although in 2D the model
would be in the universality class of the six-state clock model, leading to a Berezinskii-Kosterlitz-Thouless (BKT)
transition~\cite{jose} and no proper long-range order at finite $T$, we deliberately limit time and size
in the simulation so as to avert the undesired BKT behaviour.}.
Models similar to the present one are routinely and successfully employed in the description of the displacive structural phase transitions of
many systems, notably the very well known ferroelectric and distortive ones in the perovskites~\cite{cowley}.
In bulk simulations which we carry out first (details in Supplemental Material) the substrate equilibrium structural transition is identified
by the vanishing (near $k_B T_c = 0.075 U$) of the order parameter $\boldsymbol{\delta}(T)$, the peak
of susceptibility components $\chi_{\alpha\beta}= -(\langle u_{\alpha}u_{\beta}\rangle-\langle u_{\alpha}\rangle\langle u_{\beta}\rangle)/K_B T$
and, slightly shifted due to finite size effects, the specific heat peak $C_V=(\langle E^2\rangle-\langle E\rangle^2)/(K_B T^2 N_x N_z)$,
where $E$ is the internal energy. Below $T_c$, symmetry between the six valleys is broken
and one of the six prevails. Just above $T_c$, symmetry is thermally restored
and the distortion of each site, though instantaneously still present, is randomly distributed between all six
valleys. Near $T_c$ the system develops long correlations comparable with the simulation cell size,
and its dynamics becomes correspondingly slow, as expected at a second order phase transition.
In subsequent frictional simulations, the PBC along $z$ are removed generating two free surfaces. The point ``tip'' of
mass $M=500 m$ and coordinate $(X(t),Z(t))$, interacting with the substrate atoms via a LJ potential of depth
$V \sim 0.6U$, is dragged over one surface at constant velocity $v_0=5\cdot 10^{-3}\sqrt{U/m}$ through
a spring of constant $k=5 U/a^2$, representing the cantilever lateral stiffness. A load force $F_0$ (typically of order
$1$ in units of $U/a$) is applied along $z$ to press the tip onto the substrate.
The overall equations of motion $m \ddot{X}=-\frac{d U_{LJ}}{d X}-k(X(t)-v_0 t)$ and $m \ddot{Z}=-\frac{d U_{LJ}}{d Z}-F_0 $
are integrated with a velocity-Verlet algorithm with a time step $\Delta t=5\cdot 10^{-3} \sqrt{ma^2/U}$.
The frictional Joule heat is removed by a standard Langevin thermostat endowed with a viscous term
$-m \gamma \dot{\mathbf{r}}$ and a corresponding random noise, both attached to the slab bottom layer only
($\gamma = 10$ optimally mimics a semi-infinite substrate at an equilibrium bath temperature $T$~\cite{benassi}).
Simulation times are long enough for meaningful averages (no less than $50$ stick-slip
events required), but short enough to avert undesired order parameter destruction due either to
small size, or to the nucleation of defects heralding the onset of a BKT state. The spring elongation
$F_x(t)=-k(X(t)-v_0 t)$ measures the instantaneous friction force.
Parameters of each frictional simulation are thus a) the temperature $T$; b) the overall substrate
order parameter valley $\lambda$=1,...,6 for $T<T_c$; c) the load $F_0$; d) the average tip sliding speed $v_0$;
e) the tip effective parameters such as mass $M$ and stiffness $k$.
Among these parameters, velocity is the least critical since stick-slip is known to yields a
nearly speed independent friction coefficient~\cite{gnecco}. To cut computational costs, we generally adopt
a rather large speed ($v_0=0.005 \sqrt{U/m}$) -- except
when good quality stick-slip details are needed, requiring slower motion. Two tip mass values,
$M = 500 m$ (results shown here) or $5m$ (shown in Supplemental Material), amounting to a
factor $10$ in cantilever frequency $\sqrt{k/M}$ again gave rather similar results.
The dependence of friction upon the other parameters will be described next.

Primarily, we examine the lateral spring force $F_x(t)$, which exhibits for low speed a classic monoatomic stick-slip behaviour
(see Supplemental Material), close to that observed in realistic AFM nanofriction~\cite{mate,gnecco}. Its time average
measures the dynamic friction and the corresponding friction coefficient $\mu \equiv \langle F_x(t) \rangle /F_0$.
\begin{figure}
\centering
\includegraphics[width=8.5cm,angle=0]{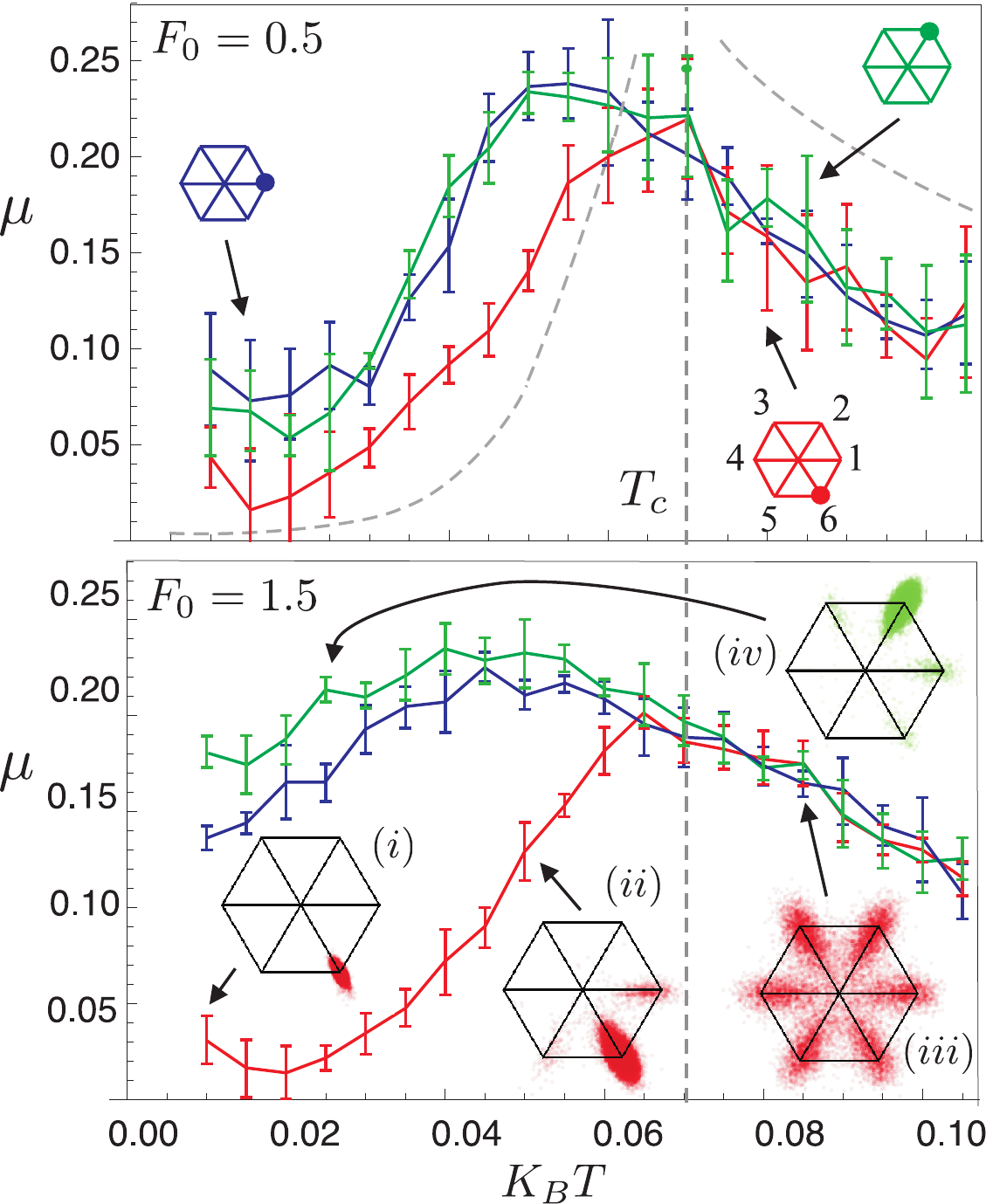}
\caption{Friction coefficient $\mu=\langle F_x \rangle/F_0$ as a function of temperature for different
substrate order parameter directions $\lambda$ (correspondence as in upper panel).
Hexagon occupancies in the lower panel illustrate the probability of the substrate ``target atom'' to be found in the
six valleys of the on-site potential after ten stick-slip events at various temperatures
above and below $T_c$ and for different distortion directions $\lambda$. Upper and lower panels refer to two distinct load values.
Dashed curve: bulk order parameter static susceptibility $\frac{\chi_{zz} (\omega =0)}{a_0^2}$.}
\label{figura2}
\end{figure}
The overall behavior obtained for friction coefficient versus temperature, order parameter valley,
and load -- our main result -- is presented in Fig.~\ref{figura2}.
First (and not surprising), the phase-transition-induced, temperature nonmonotonic stick-slip
friction is confirmed, prominent here over the (temperature-monotonic) background friction.
The friction coefficient broadly peaks in the neighborhood of the substrate phase transition, where friction
rises substantially higher than at low temperature, to descend again at higher temperatures beyond $T_c$. Second,
there is a clear dependence of friction upon the order parameter {\it valley} in the substrate. That dependence,
well visible for light loads at $T < T_c$, grows further with load (compare upper and lower panels in Fig.\ref{figura2}).
At large load and low temperature, the friction coefficient differs between valleys $\lambda =1$ (or $4$), $2$ (or $3$), and
$\lambda=6$ (or $5$) by almost an order of magnitude. We stress that the friction nonmonotony versus $T$
is in this model of a totally different nature from that recently demonstrated for multiple stick-slip regimes of
motion~\cite{tshiprut}, or for multi-contact sliders~\cite{jansen}.

So far the simulation results. Is there a linear response theory that may explain them? Strictly speaking the
answer is negative, because stick-slip is a violent, nonlinear, non-uniform perturbation. In spite of that,
it is instructive to compare the friction simulation data with a different kind of linear response theory.
If the slider's speed, jerky because of stick-slip, could be crudely replaced with a large uniform speed,
itself unperturbed by the frictional processes, then a standard ``golden rule'' linear response
could be invoked, predicting an average dissipated power
\begin{equation}
\langle P \rangle \propto \sum_\mathbf{k} \omega_\mathbf{k} \vert V_\mathbf{k}\vert^2
 \chi''(\mathbf{k},\omega_\mathbf{k}) \;,
\end{equation}
where $V_\mathbf{k}$ is the coefficient of the Fourier expansion of the slider-substrate potential,
$\omega_k = \mathbf{k}\cdot \mathbf{v}$, and $\chi''(\mathbf{k},\omega_\mathbf{k})$ is the imaginary part
of the (semi-infinite) substrate density-density correlation function, as in electron energy loss. ~\cite{pines}.
Since $\mathbf{k}\cdot \mathbf{v}$ is a low frequency, a surge of friction near $T_c$ is expected, in connection
with the increased density of low frequency modes associated with softening of the displacive mode, and eventually
with ``central peak'' diffusive excitations in the critical regime~\cite{cowley}.
Without attempting to extract $\chi''(\mathbf{k},\omega_\mathbf{k}) $ from simulations, we note
that since by Kramers-Kronig relations $\chi'(\mathbf{k} =0, \omega =0 ) = (2/\pi) \int_0^{\infty} \frac{ \chi''(0 ,\omega)}{\omega} d\omega $,
the T-dependent peak of $\chi' $ at $T_c$, while surely not identical to that of $\chi''(\mathbf{k},\omega_\mathbf{k}) $,
should similarly accompany the peaking dissipation in this approximation. Results in Fig.~\ref{figura2} indeed qualitatively confirm
a close similarity in the T-dependence of simulated stick-slip friction with that independently obtained for the bulk susceptibility.

Further insight in the order parameter valley dependence of friction can be
obtained by inspecting the system's dynamics. The slider imparts mechanical kicks to nearby substrate atoms
(details in Supplemental Material) following which, energy is transmitted to the substrate and degraded as Joule heat
(movies provided in Supplemental Material). When $T$ is low and the load is light the kicked substrate atoms
vibrate moderately, harmonically, and mainly radially along the same potential valley, see diagram (i) of Fig.~\ref{figura2},
resulting in very low friction. As the load increases, still at low temperatures, the sliding tip causes local order parameter
flips -- jumps between valleys $\lambda \rightarrow \lambda'$ -- of near-tip atoms in the substrate.
The work spent by the tip to actuate this local flip is never returned to the tip; thus an increased flip rate reflects in an
increased friction coefficient which is seen well below $T_c$.

As temperature is raised, spontaneous thermal flips of order parameter proliferate in the substrate, eventually
exploding critically near $T_c$. This is in correspondence with a surge of susceptibility, and to a drop of the free energy
barrier for the tip to cause additional sliding-induced flips; so while their number also proliferates, see Supplemental Material,
the friction rises to a critical maximum. (It should be noted however that some additional role in friction near
$T_c$ will be played by the muted propagation conditions of Joule phonons
injected into the substrate, where propagation may be impeded by critical fluctuations).
Well above $T_c$ finally all substrate atoms spontaneously and frequently jump over the six valleys
(diagram (iii) Fig.~\ref{figura2}), offering a diminishing probabilty for the slider's kicks to do work, and
friction gradually declines.
The efficiency of stick-slip in causing an order parameter flip well below $T_c$ is clearly not the same for different
valleys $\lambda$. Fig.~5 in Supplemental Material shows that
flips between valleys are more abundant when the substrate is initially polarized in $\lambda=2$ (or 1, 3, 4)
than those with $\lambda=6$ (or 5). The force exerted by the slider at slip is mostly downward oriented,
thus valleys $\lambda=2,3$ can be kicked to $\lambda'=1,4$ (or even $\lambda'=5,6$);
valleys $\lambda=1,4$ can be kicked to $\lambda'=5,6$; but valleys $\lambda=5,6$ cannot be kicked
anywheres. The frictional differences between order parameter directions become, we find, even larger
when the load is raised as shown in Fig.~\ref{figura2} lower panel. All observations remain essentially the
same for a lighter mass tip, see Supplemental Material. Although schematic, the valley-specific efficiency
difference in dissipation just demonstrated anticipates a general mechanism for the stick-slip friction dependence on the
detailed domain orientation of the substrate order parameter, and a source of AFM frictional contrast between
different domains.

\begin{figure}
\centering
\includegraphics[width=8.5cm,angle=0]{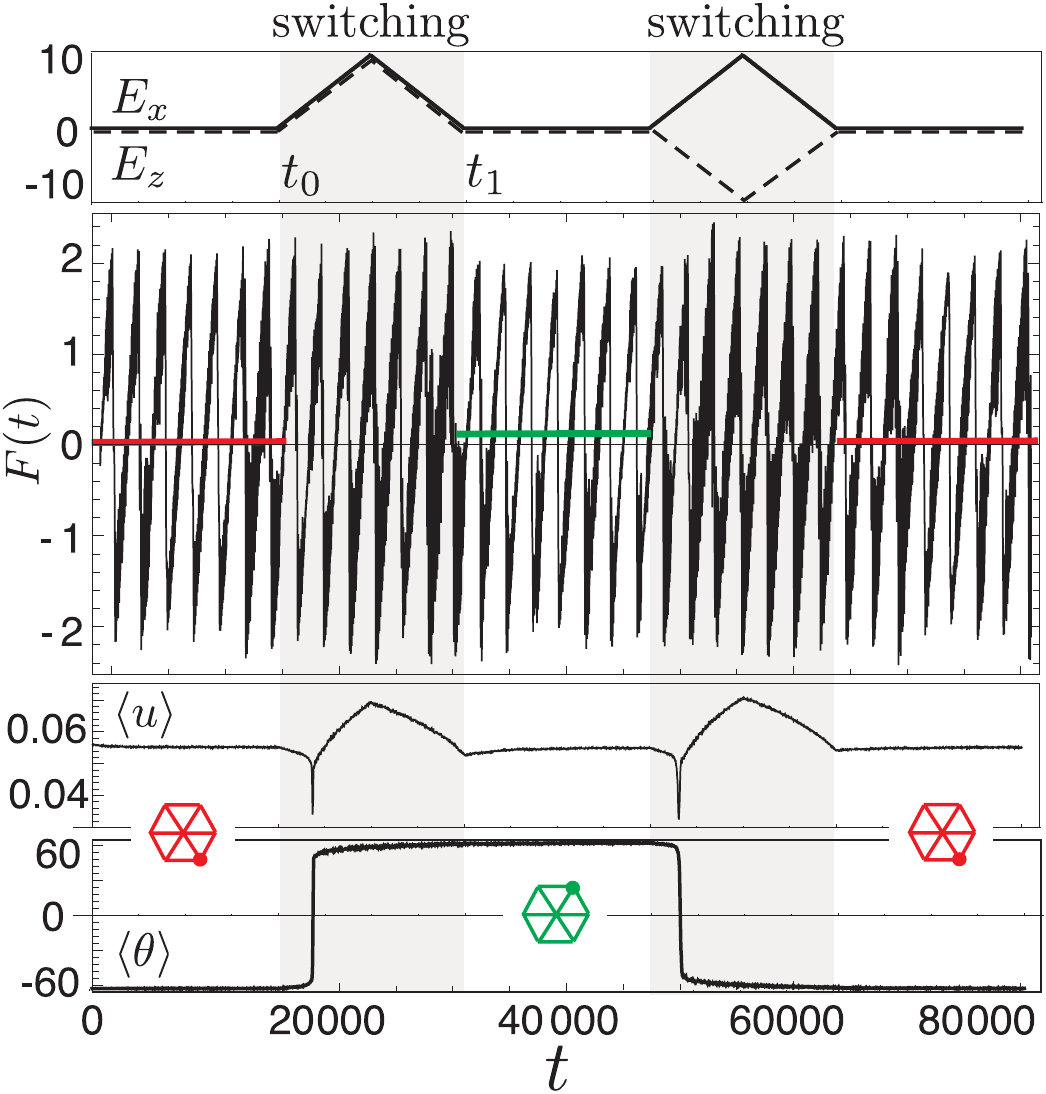}
\caption{Control of atomic stick-slip upon application of an external field, switching the overall distortion
direction from $\lambda=6$ to $\lambda'=2$. Upper panel, applied external field versus time; large panel,
stick-slip friction force; red and green lines, average friction force. Note the ``braking'' effect.
Lower panels, magnitude and orientation of the substrate order parameter. Simulations performed
at $k_BT=0.025 U$ with $F_0=1.5$ and $v_0=5\times 10^{-4}$.}
\label{figura4}
\end{figure}

This observation suggests a possible use as a means to control friction, exemplified by the simulation
of Fig.~\ref{figura4}. Start out with the substrate polarized in valley $\lambda=6$,
where the low temperature stick-slip friction is small. At time $t=t_0$, an external field
$\mathbf{E}=[E_x(t),E_z(t)]$, coupling to the order parameter in the form $\mathbf{u}\cdot \mathbf{E}$, is
turned on until $t=t_1$, when it is turned off. (For a ferroelectric substrate, $\mathbf{E}$ is an electric field;
for other ferrodistortive systems, it could be for example a uniaxial deformation). For sufficiently large field,
the substrate overall distortion switches from valley $\lambda=6$ to $\lambda'=2$, and the friction
correspondingly jumps upwards. Upon subsequent application of a restoring field $\mathbf{E}=[E_x(t),-E_z(t)]$,
friction reverts back to low.


In summary, we have explored the behaviour of the order-parameter related friction qualitatively expected for
stick-slip sliding over a structurally 'live" substrate. While quantitatively model-dependent, we obtained
answers to our three basic questions (i)-(iii) which by all signs are of wider validity. The relative magnitude
and detectability of the order parameter-related frictional effects are by necessity system-dependent
and hard to predict. On the other hand the realm of solids exhibiting (nearly) continuous structural transitions
is huge~\cite{lines_glass}. The domain contrast seen on ferroelectric BaTiO~\cite{eng2} and on ferroelastics
such as gadolinium molybdate (GMO)~\cite{czajka} and RbAlF$_4$~\cite{bulou} could be pursued by temperature
studies close to the ferro-para transitions. Antiferrodistorters like SrTiO$_3$ ($T_c \sim 105^{\circ}K$)~\cite{muller},
and antiferroelectrics such as KMnF$_3$ ($T_c \sim 187^{\circ}K$)~\cite{minkiewicz} would also be of interest.
The choice between these or other materials will largely be dictated by experimental considerations; and so will
their potential use towards realistic control over dry friction.

{\bf Acknowledgments -} This work is part of Eurocores Projects FANAS/AFRI sponsored by the Italian Research Council (CNR).
It is also sponsored in part by the Italian Ministry of University and Research, through PRIN/COFIN contracts 20087NX9Y7 and 2008Y2P573.
Discussions with H. Hug and D. Passerone are gratefully acknowledged.



\begin{thebibliography}{21}
\expandafter\ifx\csname natexlab\endcsname\relax\def\natexlab#1{#1}\fi
\expandafter\ifx\csname bibnamefont\endcsname\relax
  \def\bibnamefont#1{#1}\fi
\expandafter\ifx\csname bibfnamefont\endcsname\relax
  \def\bibfnamefont#1{#1}\fi
\expandafter\ifx\csname citenamefont\endcsname\relax
  \def\citenamefont#1{#1}\fi
\expandafter\ifx\csname url\endcsname\relax
  \def\url#1{\texttt{#1}}\fi
\expandafter\ifx\csname urlprefix\endcsname\relax\def\urlprefix{URL }\fi
\providecommand{\bibinfo}[2]{#2}
\providecommand{\eprint}[2][]{\url{#2}}

\bibitem[{\citenamefont{Persson}(1998)}]{persson_book}
\bibinfo{author}{\bibfnamefont{B.~N.} \bibnamefont{Persson}},
  \emph{\bibinfo{title}{Sliding Friction}}
  (\bibinfo{publisher}{Springer-Verlag}, \bibinfo{address}{Berlin, Germany},
  \bibinfo{year}{1998}).

\bibitem[{\citenamefont{Ala-Nissila et~al.}(1992)\citenamefont{Ala-Nissila,
  Han, and Ying}}]{ying}
\bibinfo{author}{\bibfnamefont{T.}~\bibnamefont{Ala-Nissila}},
  \bibinfo{author}{\bibfnamefont{W.~K.} \bibnamefont{Han}}, \bibnamefont{and}
  \bibinfo{author}{\bibfnamefont{S.~C.} \bibnamefont{Ying}},
  \bibinfo{journal}{Phys. Rev. Lett.} \textbf{\bibinfo{volume}{68}},
  \bibinfo{pages}{1866} (\bibinfo{year}{1992}).

\bibitem[{\citenamefont{Ala-Nissila et~al.}(2002)\citenamefont{Ala-Nissila,
  Ferrando, and Ying}}]{ying2}
\bibinfo{author}{\bibfnamefont{T.}~\bibnamefont{Ala-Nissila}},
  \bibinfo{author}{\bibfnamefont{R.}~\bibnamefont{Ferrando}}, \bibnamefont{and}
  \bibinfo{author}{\bibfnamefont{S.~C.} \bibnamefont{Ying}},
  \bibinfo{journal}{Advances in Physics} \textbf{\bibinfo{volume}{51}},
  \bibinfo{pages}{949} (\bibinfo{year}{2002}).

\bibitem[{\citenamefont{Prestipino et~al.}(1995)\citenamefont{Prestipino,
  Santoro, and Tosatti}}]{prestipino}
\bibinfo{author}{\bibfnamefont{S.}~\bibnamefont{Prestipino}},
  \bibinfo{author}{\bibfnamefont{G.}~\bibnamefont{Santoro}}, \bibnamefont{and}
  \bibinfo{author}{\bibfnamefont{E.}~\bibnamefont{Tosatti}},
  \bibinfo{journal}{Phys. Rev. Lett.} \textbf{\bibinfo{volume}{75}},
  \bibinfo{pages}{4468} (\bibinfo{year}{1995}).

\bibitem[{\citenamefont{Kisiel et~al.}(2011)\citenamefont{Kisiel, Gnecco,
  Gysin, Mariot, Rast, and Meyer}}]{kisiel}
\bibinfo{author}{\bibfnamefont{M.}~\bibnamefont{Kisiel}},
  \bibinfo{author}{\bibfnamefont{E.}~\bibnamefont{Gnecco}},
  \bibinfo{author}{\bibfnamefont{U.}~\bibnamefont{Gysin}},
  \bibinfo{author}{\bibfnamefont{L.}~\bibnamefont{Mariot}},
  \bibinfo{author}{\bibfnamefont{S.}~\bibnamefont{Rast}}, \bibnamefont{and}
  \bibinfo{author}{\bibfnamefont{E.}~\bibnamefont{Meyer}},
  \bibinfo{journal}{Nature Mat.} \textbf{\bibinfo{volume}{10}},
  \bibinfo{pages}{119} (\bibinfo{year}{2011}).

\bibitem[{\citenamefont{Dayo et~al.}(1998)\citenamefont{Dayo, Alnasrallah, and
  Krim}}]{krim}
\bibinfo{author}{\bibfnamefont{A.}~\bibnamefont{Dayo}},
  \bibinfo{author}{\bibfnamefont{W.}~\bibnamefont{Alnasrallah}},
  \bibnamefont{and} \bibinfo{author}{\bibfnamefont{J.}~\bibnamefont{Krim}},
  \bibinfo{journal}{Phys. Rev. Lett.} \textbf{\bibinfo{volume}{80}},
  \bibinfo{pages}{1690} (\bibinfo{year}{1998}).

\bibitem[{\citenamefont{Correia et~al.}(1996)\citenamefont{Correia, Massanell,
  Garcia, Levanyuk, Zlatkin, and Przeslawski}}]{correia}
\bibinfo{author}{\bibfnamefont{A.}~\bibnamefont{Correia}},
  \bibinfo{author}{\bibfnamefont{J.}~\bibnamefont{Massanell}},
  \bibinfo{author}{\bibfnamefont{N.}~\bibnamefont{Garcia}},
  \bibinfo{author}{\bibfnamefont{A.~P.} \bibnamefont{Levanyuk}},
  \bibinfo{author}{\bibfnamefont{A.}~\bibnamefont{Zlatkin}}, \bibnamefont{and}
  \bibinfo{author}{\bibfnamefont{J.}~\bibnamefont{Przeslawski}},
  \bibinfo{journal}{Appl. Phys. Lett.} \textbf{\bibinfo{volume}{68}},
  \bibinfo{pages}{2796} (\bibinfo{year}{1996}).

\bibitem[{\citenamefont{Eng et~al.}(1996)\citenamefont{Eng, Friedrich, Fousek,
  and Gunter}}]{eng2}
\bibinfo{author}{\bibfnamefont{L.}~\bibnamefont{Eng}},
  \bibinfo{author}{\bibfnamefont{M.}~\bibnamefont{Friedrich}},
  \bibinfo{author}{\bibfnamefont{J.}~\bibnamefont{Fousek}}, \bibnamefont{and}
  \bibinfo{author}{\bibfnamefont{P.}~\bibnamefont{Gunter}},
  \bibinfo{journal}{J. Vac. Sci. Technol. B} \textbf{\bibinfo{volume}{14}},
  \bibinfo{pages}{1191} (\bibinfo{year}{1996}).

\bibitem[{\citenamefont{Cowley}(1980)}]{cowley}
\bibinfo{author}{\bibfnamefont{R.~A.} \bibnamefont{Cowley}},
  \bibinfo{journal}{Advances in Physics} \textbf{\bibinfo{volume}{20}},
  \bibinfo{pages}{1} (\bibinfo{year}{1980}).

\bibitem[{\citenamefont{Lines and Glass}(1977)}]{lines_glass}
\bibinfo{author}{\bibfnamefont{M.~E.} \bibnamefont{Lines}} \bibnamefont{and}
  \bibinfo{author}{\bibfnamefont{A.~M.} \bibnamefont{Glass}},
  \emph{\bibinfo{title}{Principles and applications of ferrolectrics and
  related materials}} (\bibinfo{publisher}{Oxford University Press},
  \bibinfo{address}{Oxford}, \bibinfo{year}{1977}).

\bibitem[{\citenamefont{Benassi et~al.}(2010)\citenamefont{Benassi, Vanossi,
  Santoro, and Tosatti}}]{benassi}
\bibinfo{author}{\bibfnamefont{A.}~\bibnamefont{Benassi}},
  \bibinfo{author}{\bibfnamefont{A.}~\bibnamefont{Vanossi}},
  \bibinfo{author}{\bibfnamefont{G.~E.} \bibnamefont{Santoro}},
  \bibnamefont{and} \bibinfo{author}{\bibfnamefont{E.}~\bibnamefont{Tosatti}},
  \bibinfo{journal}{Phys. Rev. B} \textbf{\bibinfo{volume}{82}},
  \bibinfo{pages}{081401} (\bibinfo{year}{2010}).

\bibitem[{\citenamefont{Gnecco et~al.}(2000)\citenamefont{Gnecco, Bennewitz,
  Gyalog, Loppacher, Bammerlin, Meyer, and G\"untherodt}}]{gnecco}
\bibinfo{author}{\bibfnamefont{E.}~\bibnamefont{Gnecco}},
  \bibinfo{author}{\bibfnamefont{R.}~\bibnamefont{Bennewitz}},
  \bibinfo{author}{\bibfnamefont{T.}~\bibnamefont{Gyalog}},
  \bibinfo{author}{\bibfnamefont{C.}~\bibnamefont{Loppacher}},
  \bibinfo{author}{\bibfnamefont{M.}~\bibnamefont{Bammerlin}},
  \bibinfo{author}{\bibfnamefont{E.}~\bibnamefont{Meyer}}, \bibnamefont{and}
  \bibinfo{author}{\bibfnamefont{H.~J.} \bibnamefont{G\"untherodt}},
  \bibinfo{journal}{Phys. Rev. Lett.} \textbf{\bibinfo{volume}{84}},
  \bibinfo{pages}{1172} (\bibinfo{year}{2000}).

\bibitem[{\citenamefont{Mate et~al.}(1987)\citenamefont{Mate, McClelland,
  Erlandsson, and Chiang}}]{mate}
\bibinfo{author}{\bibfnamefont{C.~M.} \bibnamefont{Mate}},
  \bibinfo{author}{\bibfnamefont{G.~M.} \bibnamefont{McClelland}},
  \bibinfo{author}{\bibfnamefont{R.}~\bibnamefont{Erlandsson}},
  \bibnamefont{and} \bibinfo{author}{\bibfnamefont{S.}~\bibnamefont{Chiang}},
  \bibinfo{journal}{Phys. Rev. Lett.} \textbf{\bibinfo{volume}{59}},
  \bibinfo{pages}{1942} (\bibinfo{year}{1987}).

\bibitem[{\citenamefont{Tshiprut et~al.}(2009)\citenamefont{Tshiprut, Zelner,
  and Urbakh}}]{tshiprut}
\bibinfo{author}{\bibfnamefont{Z.}~\bibnamefont{Tshiprut}},
  \bibinfo{author}{\bibfnamefont{S.}~\bibnamefont{Zelner}}, \bibnamefont{and}
  \bibinfo{author}{\bibfnamefont{M.}~\bibnamefont{Urbakh}},
  \bibinfo{journal}{Phys. Rev. Lett.} \textbf{\bibinfo{volume}{102}},
  \bibinfo{pages}{136102} (\bibinfo{year}{2009}).

\bibitem[{\citenamefont{Barel et~al.}(2010)\citenamefont{Barel, Urbakh, Jansen,
  and Schirmeisen}}]{jansen}
\bibinfo{author}{\bibfnamefont{I.}~\bibnamefont{Barel}},
  \bibinfo{author}{\bibfnamefont{M.}~\bibnamefont{Urbakh}},
  \bibinfo{author}{\bibfnamefont{L.}~\bibnamefont{Jansen}}, \bibnamefont{and}
  \bibinfo{author}{\bibfnamefont{A.}~\bibnamefont{Schirmeisen}},
  \bibinfo{journal}{Phys. Rev. Lett.} \textbf{\bibinfo{volume}{104}},
  \bibinfo{pages}{066104} (\bibinfo{year}{2010}).

\bibitem[{\citenamefont{Pines}(1999)}]{pines}
\bibinfo{author}{\bibfnamefont{D.}~\bibnamefont{Pines}},
  \emph{\bibinfo{title}{Elementary excitations in solids}}
  (\bibinfo{publisher}{Perseus Books},
  \bibinfo{address}{Reading,Massachusetts}, \bibinfo{year}{1999}).

\bibitem[{\citenamefont{Czajka et~al.}(2000)\citenamefont{Czajka, Mielcarek,
  Mroz, Szuba, Kasuya, and Kaszczyszyn}}]{czajka}
\bibinfo{author}{\bibfnamefont{R.}~\bibnamefont{Czajka}},
  \bibinfo{author}{\bibfnamefont{S.}~\bibnamefont{Mielcarek}},
  \bibinfo{author}{\bibfnamefont{B.}~\bibnamefont{Mroz}},
  \bibinfo{author}{\bibfnamefont{S.}~\bibnamefont{Szuba}},
  \bibinfo{author}{\bibfnamefont{A.}~\bibnamefont{Kasuya}}, \bibnamefont{and}
  \bibinfo{author}{\bibfnamefont{S.}~\bibnamefont{Kaszczyszyn}},
  \bibinfo{journal}{Wear} \textbf{\bibinfo{volume}{238}}, \bibinfo{pages}{34}
  (\bibinfo{year}{2000}).

\bibitem[{\citenamefont{Bulou and Nouet}(1982)}]{bulou}
\bibinfo{author}{\bibfnamefont{A.}~\bibnamefont{Bulou}} \bibnamefont{and}
  \bibinfo{author}{\bibfnamefont{J.}~\bibnamefont{Nouet}}, \bibinfo{journal}{J.
  Phys. C.: Solid State Phys.} \textbf{\bibinfo{volume}{15}},
  \bibinfo{pages}{183} (\bibinfo{year}{1982}).

\bibitem[{\citenamefont{M\"{u}ller and Berlinger}(1971)}]{muller}
\bibinfo{author}{\bibfnamefont{K.~A.} \bibnamefont{M\"{u}ller}}
  \bibnamefont{and}
  \bibinfo{author}{\bibfnamefont{W.}~\bibnamefont{Berlinger}},
  \bibinfo{journal}{Phys. Rev. Lett.} \textbf{\bibinfo{volume}{26}},
  \bibinfo{pages}{13} (\bibinfo{year}{1971}).

\bibitem[{\citenamefont{Minkiewi and Shirane}(1969)}]{minkiewicz}
\bibinfo{author}{\bibfnamefont{V.~J.} \bibnamefont{Minkiewi}} \bibnamefont{and}
  \bibinfo{author}{\bibfnamefont{G.}~\bibnamefont{Shirane}},
  \bibinfo{journal}{J. Phys. Soc. Japan} \textbf{\bibinfo{volume}{26}},
  \bibinfo{pages}{674} (\bibinfo{year}{1969}).

\bibitem[{\citenamefont{Jos\'e et~al.}(1977)\citenamefont{Jos\'e, Kadanoff,
  Kirkpatrick, and Nelson}}]{jose}
\bibinfo{author}{\bibfnamefont{J.~V.} \bibnamefont{Jos\'e}},
  \bibinfo{author}{\bibfnamefont{L.~P.} \bibnamefont{Kadanoff}},
  \bibinfo{author}{\bibfnamefont{S.}~\bibnamefont{Kirkpatrick}},
  \bibnamefont{and} \bibinfo{author}{\bibfnamefont{D.~R.}
  \bibnamefont{Nelson}}, \bibinfo{journal}{Phys. Rev. B}
  \textbf{\bibinfo{volume}{16}}, \bibinfo{pages}{1217} (\bibinfo{year}{1977}).

\end{thebibliography}
\end{document}